\documentclass[aps,prd,12pt,a4paper,eqsecnum,nofootinbib,oneside,superscriptaddress]{revtex4-1}
\pdfoutput = 1

\usepackage{amsmath,amssymb,amsfonts,color}
\usepackage{tensor,slashed,paralist,cases,mathrsfs}
\usepackage{float,cancel,xcolor}
\usepackage{graphicx}
\usepackage{dcolumn}
\usepackage{bm}

\usepackage[colorlinks=true,urlcolor=blue,linkcolor=blue,citecolor=blue,linktocpage=true]{hyperref}

\def\be{\begin{eqnarray}}
\def\en{\end{eqnarray}}
\def\non{\nonumber}
\def\la{\langle}
\def\ra{\rangle}

\def\vp{\varepsilon}
\def\drho{\bar\rho}
\def\deta{\bar\eta}
\def\vma{{_{V-A}}}
\def\vpa{{_{V+A}}}
\def\J{{J/\psi}}
\def\ov{\overline}
\def\Lqcd{{\Lambda_{\rm QCD}}}
\def\pr{{\sl Phys. Rev.}~}
\def\prl{{\sl Phys. Rev. Lett.}~}

\def\np{{\sl Nucl. Phys.}~}

\def\lsim{ {\ \lower-1.2pt\vbox{\hbox{\rlap{$<$}\lower5pt\vbox{\hbox{$\sim$}
}}}\ } }
\def\gsim{ {\ \lower-1.2pt\vbox{\hbox{\rlap{$>$}\lower5pt\vbox{\hbox{$\sim$}
}}}\ } }

\begin{document}

\font\el=cmbx10 scaled \magstep2{\obeylines \hfill December, 2000}

\vspace*{1.5em}

\title{$B\to J/\psi K$ Decays in QCD Factorization}

\author{Hai-Yang Cheng}
\affiliation{Institute of Physics, Academia Sinica,Taipei, Taiwan 115, Republic of China}
\affiliation{Physics Department, Brookhaven National Laboratory, Upton, New York 11973}

\author{Kwei-Chou Yang}
\affiliation{Department of Physics, Chung Yuan Christian University, Chung-Li, Taiwan 320, Republic of China}


\begin{abstract}
\medskip
{The hadronic decays $B\to \J K(K^*)$ are interesting
because experimentally they are the only color-suppressed modes
which have been measured, and theoretically they are calculable by
QCD factorization even the emitted meson $\J$ is heavy. We analyze
the decay $B\to \J K$ within the framework of QCD factorization.
We show explicitly the scale and $\gamma_5$-scheme independence of
decay amplitudes and infrared safety of nonfactorizable
corrections at twist-2 order. Leading-twist contributions from the
light-cone distribution amplitudes (LCDAs) of the mesons are too
small to accommodate the data; the nonfactorizable corrections to
naive factorization are small. We study the twist-3 effects due to
the kaon and find that the coefficient $a_2(\J K)$ is largely
enhanced by the nonfactorizable spectator interactions arising
from the twist-3 kaon LCDA $\phi^K_\sigma$, which are formally
power-suppressed but chirally, logarithmically and kinematically
enhanced. Therefore, factorization breaks down at twist-3 order.
Higher-twist effects of $\J$ are briefly discussed. Our result
also resolves the long-standing sign ambiguity of $a_2(\J K)$,
which turns out to be positive for its real part.
}

\end{abstract}
\maketitle
\newpage

\section{Introduction}
There are several reasons why the decays $B\to \J K$ and $\J K^*$
are of great interest. Experimentally, they are the only color
suppressed modes in hadronic $B$ decays that have been measured so
far. These decays receive large nonfactorizable corrections, or
equivalently, the conventional parameter $a_2(\J K)$ is large, of
order $0.20-0.30$ \cite{a12}. In principle, the magnitude of
analogous $a_2$ also can be extracted directly from the decays
$\ov B^0\to D^{(*)0}\pi^0(\rho^0)$ and indirectly from the data of
$B^-\to D^{(*)}\pi(\rho)$ and $\ov B^0\to D^{(*)}\pi(\rho)$.
However, the former color-suppressed decay modes of the neutral
$B$ meson are not yet measured. Besides the form factors, the
extraction of $a_2$ from $B\to D^{(*)}\pi(\rho)$ depends on the
unknown decay constants $f_D$ and $f_{D^*}$. On the contrary, the
decay constant $f_{J/\psi}$ is well determined and the quality of
the data for $B\to J/\psi K^{(*)}$ has been significantly improved
over past years.

From the theoretical point of view, the prominent question is how
to calculate the parameter $a_2(\J K)$. In the literature, this
decay mode has been calculated using QCD sum rules and the hard
scattering approach. Khodjamirian and R\"uckl \cite{Ruckl} have
applied light-cone sum rules to study the nonfactorizable effects
in $B\to\J K_S$ and concluded that $a_2(\J K)$ is negative,
whereas Li and Yeh \cite{LY} found a positive $a_2(\J K)$ based on
the perturbative QCD hard scattering approach. Therefore, there is
a sign ambiguity for $a_2(\J K)$. Although it has been argued that
a negative $a_2(\J K)$ is very unlikely for several reasons
\cite{a12}, it is important to have an independent theory
calculation to clarify the sign issue. The QCD-improved
factorization approach advocated recently by Beneke, Buchalla,
Neubert and Sachrajda \cite{BBNS1} is suitable for this purpose.
In this approach, nonfactorizable effects in $B\to M_1M_2$ with
recoiled $M_1$ and emitted light meson $M_2$ are calculable since
only hard interactions between the $(BM_1)$ system and $M_2$
survive in the heavy quark limit.

The aforementioned QCD factorization method is no longer
applicable if the emitted meson is heavy. For example, since the
$D^0$ meson produced in  $\bar B^0\to \pi^0 D^0$ decay is not a
compact object with small transverse extension, it will interact
with the $(B\pi)$ system in the presence of soft interactions. In
principle, the parameter $a_2(\pi D)$ cannot be calculated using
QCD factorization, though it has been roughly estimated in
\cite{BBNS1} by treating the charmed meson as a light meson, which
is certainly a dubious approximation. Fortunately, the QCD
factorization approach can be applied to $B\to\J K$ decay since
the transverse size of $\J$ becomes small in the heavy quark
limit. However, a recent study by Chay and Kim \cite{Chay}
indicated that while the factorization method is applicable to
$B\to \J K$, the leading-twist contributions are too small to
explain the data.

In general, power corrections to QCD factorization are suppressed
by a factor of $\Lqcd/m_b$. However, as shown in \cite{BBNS2} and
\cite{BBNS1}, there exist power corrections which are chirally
enhanced. Moreover, there are also some power-suppressed terms at
twist-3 order which involve the logarithmically infrared
divergence, indicating the dominance of soft gluon exchange. That
is, there are twist-3 power corrections which are chirally and
logarithmically enhanced. In the present paper, we find that a
twist-3 light-cone distribution amplitude of the kaon can lead to
a large spectator interaction, which allows to alleviate the
discrepancy between theory and experiment for $B\to\J K$.

The purpose of this work on $B\to \J K$  within the framework of
QCD factorization is twofold. First of all, we perform similar
leading-twist analysis as in \cite{Chay}.  Second, we investigate
higher-twist effects on $a_2(\J K)$ and show that factorization
breaks down. The rest of this paper is organized as follows. In
Sec. II, we review the QCD factorization approach and introduce
twist-2 light-cone distribution amplitudes of mesons. Then we
proceed to compute vertex and spectator corrections to $B\to\J K$.
In Sec. III we discuss the corrections at twist-3 order and give
numerical results and discussions in Sec. IV. The conclusion is
given in Sec. V.

\section{$B\to J/\psi K$ at leading-twist order }
\subsection{QCD factorization}
Consider the hadronic decay $B\to M_1M_2$ with $M_1$ being
recoiled and $M_2$, which is a light meson or a quarkonium, being
emitted, it has been shown that the transition matrix element of
an operator $O$ in the effective weak Hamiltonian valid up to
corrections of order $\Lqcd/m_b$ is schematically given by
\cite{BBNS1}
\be
\la M_1M_2|O_i|B\ra &=& \la M_1|j_1|B\ra\la
M_2|j_2|0\ra\left[1+\sum r_n\alpha_s^n+{\cal O}({\Lqcd\over
m_b})\right]  \non \\ &=& \sum_jF_j^{BM_1}(m_2^2)\int^1_0 du\,
T_{ij}^I(u)\phi_{M_2}(u) \non \\ && +\int^1_0 d\xi \,du\,dv
\,T^{II}(\xi,u,v)\phi_B(\xi)\phi_{M_1}(v)\phi_{M_2}(u),
\label{qcdf}
\en
where $F^{BM_1}$ is a $B-M_1$ transition form factor, $\phi_M$ is
the light-cone distribution amplitude, and $T^I,~T^{II}$ are
perturbatively calculable hard scattering kernels. In the naive
factorization approach, $T^I$ is independent of $u$ as it is
nothing but the meson decay constant. However, large momentum
transfer to $M_2$ conveyed by hard gluon exchange implies a
nontrivial convolution with the distribution amplitude
$\phi_{M_2}$. The second hard scattering function $T^{II}$, which
describes hard spectator interactions, survives in the heavy quark
limit when both $M_1$ and $M_2$ are light or when $M_1$ is light
and $M_2$ is a quarkonium \cite{BBNS1}.

The factorization formula (\ref{qcdf}) implies that naive
factorization is recovered in the $m_b\to\infty$ limit and in the
absence of QCD corrections. The radiative corrections to naive
factorization are calculable since soft gluon interactions between
the $(BM_1)$ system and the $M_2$ meson are power-suppressed in
the heavy quark limit. Consequently, the nonfactorizable
contributions to naive factorization are actually amenable in the
infinite quark mass limit. The QCD factorization formula is not
applicable to the decay, for example, $\bar B^0\to \pi^0 D^0$
where the emitted meson $D^0$ is heavy so that it is neither small
( with size of order $1/\Lqcd$) nor fast and cannot be decoupled
from the $(B\pi)$ system. This is also ascribed to the fact the
soft interaction between $(B\pi)$ and the $c$ quark of the $D^0$
meson is not compensated by that between $(B\pi)$ and the light
spectator quark of the charmed meson.

In Sec. II.C we show that the QCD factorization formula is still
applicable to $B\to\J K$ decay though the emitted $\J$ is heavy.
The point is that the transverse size of $\J$ becomes small [of
order $1/(m_c\alpha_s)$] in the heavy quark limit. Technically,
infrared divergences arising from the soft interactions between
the $c$ quark of $\J$ and $(BK)$ system and between the $\bar c$
quark and $(BK)$ compensate. Therefore, the nonfactorizable
contributions to $B\to\J K$ is infrared safe.

In principle, power corrections are of order $\Lqcd/m_b$ and hence
they can be neglected in the heavy quark limit. Nevertheless,
there are some power corrections which can be enormously enhanced
and hence cannot be neglected. First of all, the contributions to,
for example $B\to K\pi$, from the $(S-P)(S+P)$ penguin operators
are enhanced by the factor
\be
{2\mu_\chi\over m_b}={2m_K^2\over (m_s+m_u)m_b}={-4\la \bar
qq\ra\over m_b f_K^2}\sim 12\,{\Lqcd\over m_b}, \label{chi}
\en
which is proportional to the quark condensate. Second, it is shown
in \cite{BBNS2} that the hard spectator interaction in $B\to K\pi$
to twist-3 order has the form
\be
f_{II} = {4\pi^2\over N_c}\,{f_K f_B\over
F_1^{B\pi}(0)m^2_B}\int^1_0 {d\drho\over\drho}\,
\phi^B(\drho)\int^1_0 {d\xi\over\xi} \,\phi^K(\xi)\int^1_0
{d\deta\over \deta}\,\left[\phi^\pi(\deta)+{2\mu_\chi\over
m_b}\phi^\pi_p(\deta)\right]. \label{fIIt3}
\en
Since the twist-3 distribution amplitude $\phi^\pi_p(\deta)\approx
1$, it does not vanish at the endpoints. Consequently, the
logarithmic divergence of the $\deta$ integral implies that
$f_{II}$ is dominated by soft gluon exchange between the spectator
quark and quarks that form the emitted pion, indicating that
factorization breaks down at twist-3 order. Hence, this power
correction is chirally and logarithmically enhanced. We shall see
in Sec. III that the hard spectator interaction in $B\to\J K$
receives the same chirally enhanced infrared logarithms from
twist-3 kaon distribution amplitudes.

\subsection{Light-cone distribution amplitudes of mesons}
Consider the matrix element of nonlocal operators sandwiched
between the vacuum and the vector meson $\J$:
\be
\la\J|\bar c_\alpha^a(x)c_\beta^b(0)|0\ra &=& {\delta^{ab}\over
4N_c}\Big\{\la \J|\bar c(x)c(0)|0\ra+\gamma_5\la \J|\bar
c(x)\gamma_5c(0)|0\ra+\gamma^\mu\la \J|\bar c(x)\gamma_\mu
c(0)|0\ra \non \\ && -\gamma^\mu\gamma_5\la \J|\bar
c(x)\gamma_\mu\gamma_5c(0)|0\ra+{1\over 2}\sigma^{\mu\nu}\la
\J|\bar c(x)\sigma_{\mu\nu}c(0)|0\ra\Big\}_{\beta\alpha},
\en
where $a,~b$ are color indices, $\alpha,~\beta$ are indices for
Dirac matrices. The leading-twist light-cone distribution
amplitudes (LCDAs) of $\J$ are given by \cite{Ballv}
\be
\la\J(P,\lambda)|\bar c(x)\gamma_\mu c(0)|0\ra  &=& f_\J
m_\J\,{\vp^{*(\lambda)}\cdot x\over P\cdot x} P_\mu\int^1_0
d\xi\,e^{i\xi P\cdot x}\phi^\J_\|(\xi), \non \\
\la\J(P,\lambda)|\bar c(x)\sigma_{\mu\nu}c(0)|0\ra &=& -if_\J^T
(\vp^{*(\lambda)}_\mu P_\nu-\vp^{*(\lambda)}_\nu P_\mu)\int^1_0
d\xi\,e^{i\xi P\cdot x}\phi^\J_\bot(\xi), \label{Jwf}
\en
where $\vp^*$ is the polarization vector of $\J$, $\xi$ is the
light-cone momentum fraction of the $c$ quark in $\J$, $f_\J$ and
$f^T_\J$ are vector and tensor decay constants, respectively, but
the latter is scale dependent. The normalization conditions of the
twist-2 LCDAs are
\be
\int^1_0d\xi\,\phi^\J_\|(\xi)=\int^1_0d\xi\,\phi^\J_\bot(\xi)=1.
\en
Likewise, at the twist-2 accuracy, the only kaon LCDA which
contributes to the matrix element (the relation
$\sigma^{\mu\nu}\otimes
\sigma_{\mu\nu}=\sigma^{\mu\nu}\gamma_5\otimes
\sigma_{\mu\nu}\gamma_5$ being applied)
\be
\la K^-|\bar s_\alpha^a(0)u_\beta^b(x)|0\ra &=& {\delta^{ab}\over
4N_c}\Big\{\la K^-|\bar s(0)u(x)|0\ra+\gamma_5\la K^-|\bar
s(0)\gamma_5u(x)|0\ra+\gamma^\mu\la K^-|\bar s(0)\gamma_\mu
u(x)|0\ra \non \\ && -\gamma^\mu\gamma_5\la K^-|\bar
s(0)\gamma_\mu\gamma_5u(x)|0\ra+{1\over
2}\sigma^{\mu\nu}\gamma_5\la K^-|\bar
s(0)\sigma_{\mu\nu}\gamma_5u(x)|0\ra\Big\}_{\beta\alpha},
\en
is given by  \cite{Ballp}
\be
\la K^-(P)|\bar s(0)\gamma_\mu\gamma_5 u(x)|0\ra =-if_K
P_\mu\int^1_0 d\deta\,e^{i\deta P\cdot x}\phi^K(\deta),
\en
where $\deta$ is the momentum fraction of the antiquark $\bar u$
in $K^-$ and $\int^1_0\phi^K(\deta)d\deta=1$.

As for the $B$ meson LCDA, we will follow \cite{BBNS1} to choose
\be
\la 0|\bar q_\alpha(x)b_\beta(0)|\bar
B(p)\ra\!\!\mid_{x_+=x_\bot=0}=-{if_B\over 4}[(p\!\!\!/
+m_B)\gamma_5]_{\beta\gamma}\int^1_0d\drho\, e^{-i\drho
p_+x_-}[\phi^B_1(\drho)+n\!\!\!/_-\phi^B_2(\drho)]_{\gamma\alpha},  \hskip0.7cm
\label{Bwf}
\en
based on the observation that the $B$ meson is described by two
scalar wave functions at the leading order in $1/m_b$, where
$\drho$ is the momentum fraction carried by the spectator quark of
the $B$ meson. In Eq. (\ref{Bwf}), $n_-=(1,0,0,-1)$ and the
normalization conditions are
\be
\int^1_0d\drho \,\phi^B_1(\drho)=1, \qquad\quad \int^1_0d\drho\,
\phi^B_2(\drho)=0.
\en

The leading-twist LCDAs of $\J$ can be expanded as \cite{Ballv},
\be
\phi^\J_\|(\xi) &=& 6\xi(1-\xi)\left(1+{3\over
2}\,a_2^\|\,[5(2\xi-1)^2-1]\right),  \non \\ \phi^\J_\bot(\xi) &=&
6\xi(1-\xi)\left(1+{3\over 2}\,a_2^\bot\,[5(2\xi-1)^2-1]\right),
\label{Jlcda}
\en
where the parameters $a_2^\|$ and $a_2^\bot$ are defined by the
matrix element of a twist-2 conformal operator with conformal spin
3 \cite{Ballv},\footnote{Since the parameters $a_2^\|$ and
$a_2^\bot$ are unknown, we will employ the asymptotic DAs
$\phi^\J_\|(\xi)=\phi^\J_\bot(\xi)=6\xi(1-\xi)$ in numerical
calculations in Sec. IV. For general discussions in Secs. II and
III, we still use Eq. (\ref{Jlcda}) for the LCDAs of $\J$.} while
twist-2 DA $\phi^K$ can be expanded in terms of Gegenbauer
polynomials $C^{3/2}_n$ \cite{Ballp}:
\be
\phi^K(\deta,\mu^2)=6\deta(1-\deta)\left(1+\sum_{n=1}^\infty
a_{2n}^K(\mu^2)C_{2n}^{3/2}(2\deta-1)\right),
\en
with the values of the Gegenbauer moments $a_n^K$ being available
from \cite{Ballp}. For the $B$ meson, we use \cite{Keum}
\be
\phi^B_1(\drho)=N_B\drho^2(1-\drho)^2{\rm exp}\left[-{1\over
2}\left({\drho m_B\over \omega_B}\right)^2\right],
\en
with $\omega_B=0.25$ GeV and $N_B$ being a normalization constant.
This $B$ meson wave function corresponds to $\lambda_B=303$ MeV
defined by $\int^1_0 d\drho\,\phi^B(\drho)/\drho\equiv
m_B/\lambda_B$. This can be understood since the $B$ meson wave
function is peaked at small $\drho$: It is of order $m_B/\Lqcd$ at
$\drho\sim \Lqcd/m_B$. Hence, the integral over
$\phi_B(\drho)/\drho$ produces an $m_B/\Lqcd$ term.

In ensuing calculations we find that the contributions from the
$\J$ LCDA $\phi^\J_\bot$ are proportional to the factor $(f_\J^T
m_c)/(f_\J m_\J)$ relative to that from $\phi^\J_\|$. Contracting
$\la \J|\bar c(0)\sigma_{\mu\nu}c(0)|0\ra$ in Eq. (2.4) with
$P^\nu$ and applying the equation of motion and the matrix element
\be
\la \J|\bar c(0)\gamma_\mu c(0)|0\ra= f_\J m_\J \vp^*_\mu,
\en
we are led to \cite{Chay}
\be
f_\J^T m_\J=2f_\J m_c.
\en
Therefore,
\be
{f_\J^T m_c\over f_\J m_\J}=2\left({m_c\over
m_\J}\right)^2=2\xi^2, \label{relation}
\en
where in the last step we have applied the on-shell relation
$\xi^2 P\!\!\!\!/^{\,2}_\J=m_c^2$ for the charmed (not
anticharmed) quark in the $\J$.

\subsection{$B\to\J K$ in QCD factorization}
The effective Hamiltonian relevant for $B\to \J K$  has the form
\be \label{hamiltonian} {\cal H}_{\rm eff} =
{G_F\over\sqrt{2}}\Bigg\{V_{cb}V_{cs}^*
\Big[c_1(\mu)O_1(\mu)+c_2(\mu)O_2(\mu)\Big]
-V_{tb}V_{ts}^*\sum^{10}_{i=3}c_i(\mu)O_i(\mu)\Bigg\}+{\rm h.c.},
\en
where
\be
 && O_1= (\bar cb)_\vma(\bar sc)_\vma,
\qquad\qquad\qquad\qquad\quad~ O_2 = (\bar sb)_\vma(\bar cc)_\vma,
\non
\\ && O_{3(5)}=(\bar sb)_\vma\sum_{q'}(\bar q'q')_{\vma(\vpa)},
\qquad  \qquad~ O_{4(6)}=(\bar s_\alpha
b_\beta)_\vma\sum_{q'}(\bar q'_\beta q'_\alpha)_{ \vma(\vpa)},
\\ && O_{7(9)}={3\over 2}(\bar sb)_\vma\sum_{q'}e_{q'}(\bar
q'q')_{\vpa(\vma)},
  \qquad~ O_{8(10)}={3\over 2}(\bar s_\alpha b_\beta)_\vma\sum_{q'}e_{q'}(\bar
q'_\beta q'_\alpha)_{\vpa(\vma)},   \non
\en
with $O_3$--$O_6$ being the QCD penguin operators,
$O_{7}$--$O_{10}$ the electroweak penguin operators, and $(\bar
q_1 q_2)\equiv\bar q_1\gamma_\mu(1-\gamma_5)q_2$.

Under naive factorization, the decay amplitude of $B\to\J K$ reads
\be
A(B\to\J
K)=\,{G_F\over\sqrt{2}}V_{cb}V_{cs}^*(a_2+a_3+a_5+a_7+a_9)f_\J
m_\J F_1^{BK}(m^2_\J)(2\vp^*\cdot p_B),  \label{naivef}
\en
where $a_{2i}=c_{2i}+(1/N_c)c_{2i-1}$ and
$a_{2i-1}=c_{2i-1}+(1/N_c)c_{2i}$ in naive factorization and the
approximation $V_{tb}V_{ts}^*\approx -V_{cb}V_{cs}^*$ has been
made. The form factor $F_1^{BK}$ is defined by
\be
\la P'(p')|V_\mu|P(p)\ra = \left(p_\mu+p'_\mu-{m_P^2-m_{P'}^2\over
q^2}\,q_ \mu\right) F_1(q^2)+{m_P^2-m_{P'}^2\over
q^2}q_\mu\,F_0(q^2).
\en
There are two serious problems with the naive factorization
approximation. First, the Wilson coefficients $c_i(\mu)$ and hence
$a_i$ are renormalization scale and $\gamma_5$-scheme dependent,
whereas the decay constants and form factors are not. Hence, the
amplitude (\ref{naivef}) is not physical. Second, nonfactorizable
effects, which play an essential role in color-suppressed modes,
are not taken into account.

The aforementioned two difficulties for naive factorization are
resolved in the QCD factorization approach in which the inclusion
of vertex corrections and hard spectator interactions (see Fig. 1)
yields\footnote{Using the constant matrices $\hat{r}_{\rm NDR}$
and $\hat{r}_{\rm HV}$ given in Eqs. (2.18) and (2.19) in
\cite{CCTY}, it is straightforward to obtain the constant terms in
Eq. (\ref{ai}) in NDR and HV $\gamma_5$-schemes.}
\be
\label{ai}
a_2 &=& c_2+{c_1\over N_c}
+{\alpha_s\over4\pi}\,{C_F\over N_c} c_1
\left[-\left(
\begin{matrix}18\cr 14\cr \end{matrix}
\right)-12\ln{\mu\over m_b}+f_I+f_{II}+{F_0^{BK}(m^2_\J)\over F_1^{BK}(m^2_\J)}g_I\right], \non \\ 
a_3 &=& c_3+{c_4\over N_c}
+{\alpha_s\over 4\pi}\,{C_F\over N_c} c_4\left[-\left(
\begin{matrix} 18\cr 14\cr\end{matrix}\right)-12\ln{\mu\over m_b}+f_I+f_{II}+{F_0^{BK}(m^2_\J)\over F_1^{BK}(m^2_\J)}g_I\right], \non \\ 
a_5 &=& c_5+{c_6\over N_c}
-{\alpha_s\over 4\pi}\,{C_F\over N_c} c_6\left[-\left(
\begin{matrix} 6\cr 18\cr\end{matrix}\right)-12\ln{\mu\over m_b}+f_I+f_{II}+{F_0^{BK}(m^2_\J)\over F_1^{BK}(m^2_\J)}g_I\right],  \\ 
a_7 &=& c_7+{c_8\over N_c}-{\alpha_s\over 4\pi}\,{C_F\over N_c} c_8\left[-\left(
\begin{matrix} 6\cr 18\cr\end{matrix}\right)-12\ln{\mu\over m_b}+f_I+f_{II}+{F_0^{BK}(m^2_\J)\over F_1^{BK}(m^2_\J)}g_I\right], \non \\ 
a_9 &=& c_9+{c_{10}\over N_c}+{\alpha_s\over 4\pi}\,{C_F\over N_c} c_{10}\left[-\left(
\begin{matrix}  18\cr 14\cr\end{matrix}\right)-12\ln{\mu\over m_b}+f_I+f_{II}+{F_0^{BK}(m^2_\J)\over F_1^{BK}(m^2_\J)}g_I\right], \non
\en
where the upper entry of the matrix is evaluated in the naive
dimension regularization (NDR) scheme and the lower entry in the
't Hooft-Veltman (HV) renormalization  scheme,
$C_F=(N_c^2-1)/(2N_c)$, and $N_c$ is the number of colors. The
hard scattering functions $f_I$ and $g_I$ arise from the vertex
corrections, Figs. 1(a)-1(d), while $f_{II}$ from the hard
spectator interactions Figs. 1(e)-1(f). Formally, the coefficients
$a_i$ are scale and $\gamma_5$-scheme independent and we will come
to this point again in Sec. IV.

The results for the hard scattering functions $f_I$ and $g_I$ are
\be
f_I &=& \int^1_0 d\xi\,\phi^\J(\xi)\Bigg\{ {2z\xi\over
1-z(1-\xi)}+\left(3-2\xi-8\xi^2\right){\ln \xi\over 1-\xi} \non
\\ &&+\left(-{3\over 1-z\xi}+{1+8\xi\over
1-z(1-\xi)}-{2z\xi\over [(1-z(1-\xi)]^2}\right)z\xi\ln z\xi \non
\\ &&+\left(3(1-z)+2z\xi-8z\xi^2+{2z^2\xi^2\over
1-z(1-\xi)}\right){\ln (1-z)-i\pi\over 1-z(1-\xi)},    \label{fI}
\en
and
\be
g_I &=& \int^1_0 d\xi\,\phi^\J(\xi)\Bigg\{{4\xi(2\xi-1)\over
(1-z)(1-\xi)}\ln \xi+{z\xi\over [1-z(1-\xi)]^2}\ln (1-z)  \non
\\ &&+ \Bigg({1\over (1-z\xi)^2}-{1\over
[1-z(1-\xi)]^2}-{8\xi\over (1-z)(1-z\xi)}  \non \\
&&+{2(1+z-2z\xi)\over (1-z)(1-z\xi)^2}\Bigg)z\xi\ln
z\xi-i\pi\,{z\xi\over [1-z(1-\xi)]^2}\Bigg\}, \label{gI}
\en
where $z\equiv m_\J^2/m_B^2$ and the assumption
$\phi^\J(\xi)=\phi^\J_\|(\xi)=\phi^\J_\bot(\xi)$ has been made. In
deriving Eqs. (\ref{fI}) and (\ref{gI}) we have applied
(\ref{relation}) and the relation $\phi^\J(\bar\xi)=\phi^\J(\xi)$,
where $\bar\xi=1-\xi$ is the light-cone momentum fraction of the
$\bar c$ quark in $\J$.

Several remarks are in order. (i) We have proved explicitly the
cancellation of infrared divergences so that the resultant
amplitude is infrared finite when summing all the diagrams in Fig.
1, a key element for the applicability of QCD factorization.
Moreover, we also show a cancellation of the infrared double
poles, i.e. $1/\epsilon_{\rm IR}^2$,  as in \cite{CLY}. As
stressed and elucidated in \cite{Chay}, in the limits $z\to 0$ and
$z\neq 0$, the amplitudes are infrared finite in both cases, but
for different reasons. (ii) The strong phases in $f_I$ and $g_I$
arise from the diagrams Figs. 1(c)-1(d) where there are hard gluon
exchanges between the outgoing $K$ and $\J$. (iii) Only the form
factor $F_1^{BK}$ contributes to the decay amplitude under naive
factorization, while  to the order $\alpha_s$ both form factors
$F_0^{BK}$ and $F_1^{BK}$ contribute. (iv) Our results for $f_I$
and $g_I$ agree with that in \cite{Chay} by noting that the form
factors $F_{0,1}^{BK}$ are related via
$F_0^{BK}(m_\J^2)/F_1^{BK}(m_\J^2)=(m_B^2-m_\J^2)/m_B^2$
\cite{Chay}. The only difference is that we treat the ratio
$(f^T_\J m_c)/(f_\J m_\J)$ as $2\xi^2$ [see Eq. (\ref{relation})],
while it is considered to be a constant in \cite{Chay}. (v) It is
easily seen that in the zero charmed quark mass limit,
\be
f_I+g_I=\int^1_0 d\xi\,\phi^\J(\xi)\left(3{1-2\xi\over
1-\xi}\ln\xi-3i\pi\right),
\en
in agreement with  \cite{BBNS1} for $B\to\pi\pi$, as it should be.

\begin{figure}[t!]
\begin{center}
\vspace{-3.5cm}
\includegraphics[height=7.5 in,width=6.5in]{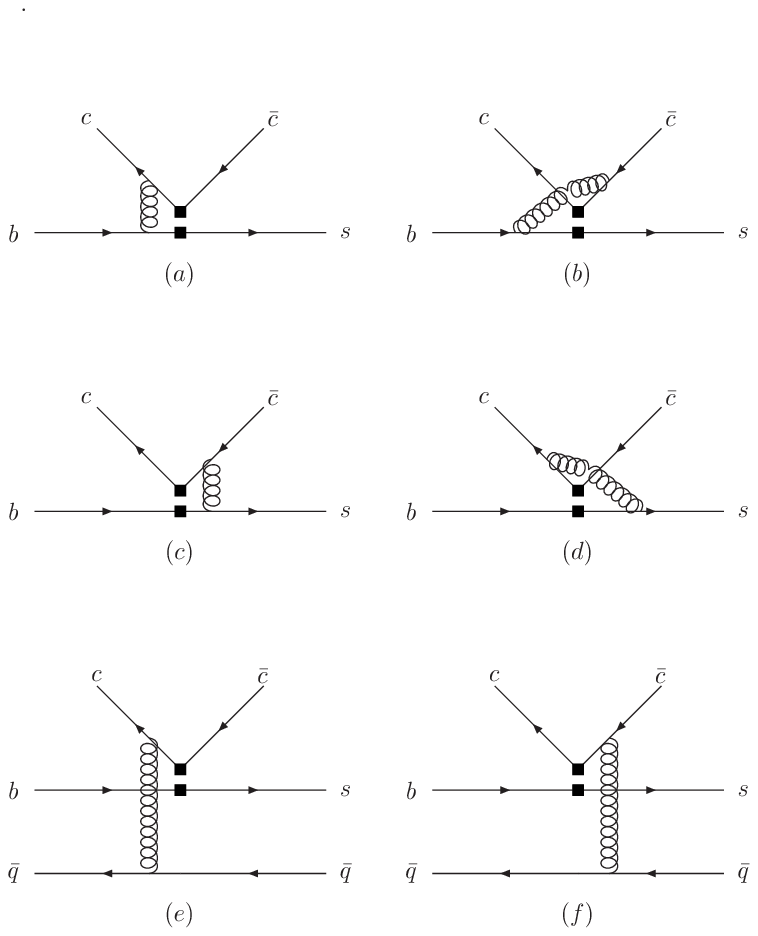}
\vspace{-5.5cm}
    \caption{{\small Vertex and spectator corrections to $B\to J/\psi K$.}}
   \label{fig:vert}
\end{center}
\end{figure}

As for the hard scattering function $f_{II}$ originating from
spectator diagrams, we write
\be
f_{II}=f_{II}^2+f_{II}^3+\cdots,
\en
where the superscript denotes the twist dimension of LCDA. To the
leading-twist order, we obtain
\be
f^2_{II} &=& {4\pi^2\over N_c}\,{f_Kf_B\over
F_1^{BK}(m^2_\J)m^2_B}\int^1_0 d\xi\, d\drho\,
d\deta\,\phi^B_1(\drho)\phi^\J(\xi)\phi^K(\deta)  \non \\ &\times
& { \drho-\deta+(\drho-2\xi+\deta)z+4\xi^2z\over
\drho(\drho-\deta+\deta
z)[(\drho-\xi)(\drho-\deta)+(\deta\drho-\deta\xi-\drho\xi)z]}.
\label{fII2}
\en
This can be further simplified by noting that $\drho\sim {\cal
O}(\Lqcd/m_b)\to 0$ in the $m_b\to\infty$ limit. Hence,
\be
f^2_{II} = {4\pi^2\over N_c}\,{f_Kf_B\over
F_1^{BK}(m^2_\J)m^2_B}\,{1\over 1-z}\int^1_0 d\drho\,
{\phi^B_1(\drho)\over \drho}\int^1_0 d\xi \,{\phi^\J(\xi)\over
\xi}\int^1_0 d\deta\, {\phi^K(\deta)\over \deta}, \label{fII2'}
\en
where the $z$ terms in the numerator cancel after the integration
over $\xi$.\footnote{Although the same expression (\ref{fII2'}) is
also given in \cite{Chay}, it is not clear to us how to achieve
the cancellation of the $z$ terms appearing in the numerator of
Eq. (\ref{fII2}) when $4\xi^2$ is replaced by $2(f_\J^T m_c)/(f_\J
m_\J)$ and treated as a constant according to \cite{Chay}.}

Before proceeding we make two remarks: (i) At a first glance, it
appears that the spectator contribution is power suppressed by
order $(\Lqcd/m_b)^2$. This is not the case. First of all, the $B$
wave function $\phi_1^B(\drho)$ is of  order $m_b/\Lqcd$ at
$\drho\sim \Lqcd/m_b$ [see also the discussion after Eq.
(\ref{Bwf})]. Second, the form factor $F_1^{BK}$ scales as $\sim
(\Lqcd/m_b)^{3/2}$, while decay constants as $f_B\sim
\Lambda^{3/2}_{\rm QCD}/m_b^{1/2}$ and $f_\pi\sim \Lqcd$
\cite{BBNS1}. It follows that $f_{II}^2$ is of the same order as
$f_I$ and $g_I$ and is not power suppressed, contrary to the claim
in \cite{Chay}. Furthermore, the former is numerically more
important than the latter, as we shall see in Sec. IV. (ii) By
inspection of (\ref{fII2}), there will be some strong phases
coming from hard spectator interactions beyond the heavy quark
limit, which are difficult to evaluate numerically, however.

\section{Higher-twist effects}
In the last section we have computed leading-twist vertex and hard
spectator corrections to naive factorization. However, we shall
see in Sec. IV that numerically the twist-2 nonfactorizable
effects are small; the predicted decay rate of $B\to\J K$ is too
small by a factor of $7\sim 10$. Therefore, it is inevitable that
higher-twist effects which are seemingly power suppressed should
play an essential role.

As shown in Sec. II.A, twist-3 power corrections to the spectator
interactions could be enormously enhanced by the chirally enhanced
infrared logarithms. One reason that the spectator diagrams could
receive large power corrections  is ascribed to the fact that the
hard gluon exchange in the spectator diagram is not hard enough.
The virtual gluon's momentum squared is
\be
k^2=(-\drho p_B+\deta p_K)^2\approx -\drho\deta m_B^2\sim -\Lqcd
m_b,
\en
which is of order 1 GeV due to the smallness of the momentum
fraction $\drho\sim \Lqcd/m_b$ carried by the spectator quark in
the $B$ meson. Therefore, we shall study higher-twist power
corrections to hard spectator interactions.
\subsection{Twist-3 LCDAs of the kaon}
We first consider the twist-3 DAs $\phi^K_p$ and $\phi^K_\sigma$
of the kaon defined in the pseudoscalar and tensor matrix elements
\cite{Ballp}:
\be
\la K^-(P)|\bar s(0)i\gamma_5 u(x)|0\ra &=& {f_K m_K^2\over
m_s+m_u}\int^1_0 d\deta\,e^{i\deta P\cdot x}\phi_p^K(\deta), \non
\\ \la K^-(P)|\bar s(0)\sigma_{\mu\nu}\gamma_5 u(x)|0\ra &=&
-{i\over 6}{f_K m_K^2\over m_s+m_u}\left[1-\left({m_s+m_u\over
m_K}\right)^2\right]  \non \\ && \times(P_\mu x_\nu-P_\nu x_\mu)
\int^1_0 d\deta \,e^{i\deta P\cdot x}\phi_\sigma^K(\bar \eta).
\label{ktw3}
\en
They can be expanded in terms of Gegenbauer polynominals
\cite{Ballp}:
\be
\phi^K_p(\deta) &=& 1+aC_2^{1/2}(\deta)+bC_4^{1/2}(\deta)+\cdots,
\non
\\ \phi^K_\sigma(\deta) &=& 6\deta(1-\deta)(1+dC_2^{3/2}(\deta)+\cdots),
\en
where the coefficients $a,~b,~d$ can be found in \cite{Ballp}.
From Eq. (\ref{ktw3}) it is clear that twist-3 DAs of pseudoscalar
mesons  are associated with a chiral enhancement factor
$\mu_\chi$. As before, $\deta$ is the light-cone momentum fraction
of the $\bar u$ quark in $K^-$. Just as the hard spectator
interactions in $B\to K\pi$ decay which receive large
power-suppressed corrections from twist-3 LCDAs of the pion (see
Sec. II.A), we find that the twist-3 kaon LCDA $\phi^K_\sigma$
contributes to spectator diagrams in $B\to\J K$ decay and yields
\be
f^3_{II} &=& {4\pi^2\over N_c}\,{f_K f_B\over
F_1^{BK}(m^2_{J/\psi})m_B^2}\,{2m_K^2\over (m_s+m_u)m_B^2}\int^1_0
d\xi\,d\drho\,d\bar\eta\,\phi^B_1(\bar\rho)\phi^{J/\psi}
(\xi)\phi^K_\sigma(\deta) \non
\\ &\times & {1\over 6}\,\left( {1\over
\xi\bar\rho\deta^2(1-z)^2}-{4z\over \drho\deta^3(1-z)^3}+{8\xi
z\over \drho\deta^3(1-z)^3}\right).
\en
Applying the twist-3 DA $\phi^K_\sigma(\deta)=6\deta(1-\deta)$, we
see that the linear divergent terms proportional to $\int^1_0
d\deta/\deta^2$ cancel after the integration over $\xi$; this
cancellation happens because the $\J$ distribution amplitude is
even in $(2\xi-1)$ [see Eq. (\ref{Jlcda})]. Consequently,
\be
f^3_{II} =\left({2\mu_\chi\over m_B}\right)\,{4\pi^2\over
N_c}\,{f_K f_B\over F_1^{BK}(m^2_{J/\psi})m_B^2} \int^1_0
{d\drho\over \drho}\,\phi^B_1(\drho)\int^1_0 {d\xi\over
\xi}\,\phi^\J(\xi)\int^1_0 {d\deta\over \deta^2}\,
{\phi^K_\sigma(\deta)\over 6(1-z)^2}, \label{fII3}
\en
with $\mu_\chi$ being defined in Eq. (\ref{chi}). Therefore, to
the twist-3 order of kaon DA,
\be
f_{II}=f_{II}^2+f_{II}^3 &=& {4\pi^2\over N_c}\,{f_K f_B\over
F_1^{BK}(m^2_{J/\psi})m_B^2}\,{1\over 1-z} \int^1_0 {d\drho\over
\drho}\,\phi^B_1(\drho)\int^1_0 {d\xi\over \xi}\,\phi^\J(\xi) \non
\\ && \times\int^1_0 {d\deta\over \deta}\left(\phi^K(\deta)+{2\mu_\chi\over
m_B}\, {1\over (1-z)}\,{\phi^K_\sigma(\deta)\over
6\deta}\right), \label{fIIt}
\en
which, apart from the constant term, agrees with Eq. (\ref{fIIt3})
in the $z\to 0$ limit.\footnote{Just as $B\to \J K$ decay, we find
that the twist-3 pion distribution amplitude that contributes
directly to the spectator interactions of $B\to K\pi$ is
$\phi^\pi_\sigma$ rather than $\phi^\pi_p$. More specifically, we
have a contribution like
$X=\int^1_0(d\deta/\deta^2)\phi^\pi_\sigma/6$. At a first glance,
it seems to be different than the last term appearing in Eq.
(\ref{fIIt3}). However, since the LCDAs $\phi_p$ and $\phi_\sigma$
are related by equations of motion \cite{BF},
$\phi_\sigma(\deta)/6=\deta(\phi_p(\deta)/2+\phi'_\sigma(\deta)/12)$,
it follows that
$X=\int^1_0(d\deta/\deta)(\phi_p(\deta)/2+\phi'_\sigma(\deta)/12)$.
This means that $X$ can be expressed in terms of $\phi_p$ and
$\phi'_\sigma$ which do not vanish at endpoints. Since
$\phi'_\sigma(\deta)=6(1-2\deta)$, $X$ is reduced to $\int^1_0
(d\deta/\deta)\phi_p(\deta)$ after the non-logarithmic term is
dropped. Nevertheless, we wish to emphasize that the logarithmic
divergence arises originally from $\phi_\sigma$ instead of
$\phi_p$.} The logarithmic divergence of the $\deta$ integral
implies that the spectator interaction is dominated by soft gluon
exchanges between the spectator quark and the charmed or
anti-charmed quark of $\J$. Hence, factorization breaks down to
twist-3 order. Following \cite{BBNS2}, we treat the divergent
integral as an unknown parameter and write
\be
X\equiv\int^1_0 {d\deta\over
\deta}=\ln\left({m_B\over\Lqcd}\right)+r, \label{logdiv}
\en
with $r$ being a complex random number. Therefore, although
$f_{II}^3$ is formally power suppressed in the heavy quark limit,
it is chirally enhanced by a factor of $(2\mu_\chi/\Lqcd)\sim 12$,
logarithmically enhanced by the infrared logarithms and
kinematically enhanced by a factor of $1/(1-z)^2$. Numerically,
nonfactorizable corrections to naive factorization are dominated
by $f_{II}^3$ (see Sec. IV).

\subsection{Higher-twist LCDAs of $\J$}
Intuitively it is expected that light mesons produced in hadronic
$B$ decays are appropriately described by LCDAs. Since $\J$ is not
light enough, it is nature to conjecture the importance of
higher-twist effects of $\J$. The twist-3 LCDAs of $\J$ are given
by \cite{Ballv}
\be
\la\J(P,\lambda)|\bar c(x)\gamma_\mu c(0)|0\ra  &=& \cdots+ f_\J
m_\J\left(\vp^{*(\lambda)}_\mu-{\vp^{*(\lambda)}\cdot x\over
P\cdot x} P_\mu\right)\int^1_0 d\xi\,e^{i\xi P\cdot
x}g_\bot^{(v)}(\xi), \non
\\ \la\J(P,\lambda)|\bar c(x)\gamma_\mu\gamma_5 c(0)|0\ra &=&
-{1\over 4}m_\J\left(f_\J-f_\J^T\,{2m_c\over m_\J}\right)   \non
\\
&&\times \epsilon_{\mu\nu\alpha\beta}\vp^{*\nu}P^\alpha
x^\beta\int^1_0 d\xi\,e^{i\xi P\cdot x}g_\bot^{(a)}(\xi),
\en
for chiral-even DAs $g_\bot^{(v)}$ and $g_\bot^{(a)}$,
\be
\la\J(P,\lambda)|\bar c(x)\sigma_{\mu\nu} c(0)|0\ra  &=& \cdots
-if^T_\J (P_\mu x_\nu-P_\nu x_\mu){\vp^{*(\lambda)}\cdot x\over
(P\cdot x)^2}m_\J^2\int^1_0 d\xi\,e^{i\xi P\cdot
x}h_\|^{(t)}(\xi), \non
\\ \la\J(P,\lambda)|\bar c(x) c(0)|0\ra &=&
i\left(f_\J-f_\J^T\,{2m_c\over m_\J}\right)(\vp^{*}\cdot
x)m_\J^2\int^1_0 d\xi\,e^{i\xi P\cdot x}h_\|^{(s)}(\xi),
\en
for chiral-odd DAs $h_\|^{(t)}$ and $h_\|^{(s)}$, where ellipses
denote twist-2 LCDAs given by (\ref{Jwf}).

The calculation of twist-3 effects for $\J$ is more complicated
since it involves four more LCDAs and some other technical
problems. Nevertheless, we can argue that the DAs $g_\bot^{(a)}$
and $h_\|^{(s)}$ are less important. From Eq. (\ref{relation}) it
follows that the term
\be
f_\J-{2m_c\over m_\J}f_\J^T=f_\J\left(1-{4m_c^2\over
m_\J^2}\right)
\en
vanishes in the heavy quark limit. Therefore, it is of order
$\Lqcd/m_b$. A full study of the twist-3 effects of $\J$ is in
progress. However, just as the $B\to K\pi$ decay discussed in Sec.
II.A, it is expected that twist-3 corrections which manifest
predominately in hard spectator interactions are dominated by the
kaon.

\section{Numerical results and discussions}
Before proceeding to the numerical results, we first demonstrate
the scale and scheme independence of the coefficients $a_i$ given
in Eq. (\ref{ai}). To show the scale independence of $a_3$, for
example, we note
\be
{d\,c_i(\mu)\over d\ln \mu }=\,{\alpha_s\over
4\pi}\,\gamma^T_{ij}c_i(\mu),
\en
where the anomalous dimension matrix $\gamma$ can be found in the
literature (see e.g. \cite{Buras96}). We find
\be
{d\over d\ln \mu}(c_3+{c_4\over 3})=\,12\,{\alpha_s\over
4\pi}\,{C_F\over N_c} c_4
\en
and hence $da_3/d\ln\mu=0$ to ${\cal O}(\alpha_s^2)$, as it should
be.\footnote{Empirically one will find that there is a slight
scale dependence of $a_i$ owing to the fact that not all leading
logarithmic corrections to $a_i$ to all orders in $\alpha_s$ are
included in Eq. (\ref{ai}).} As for the $\gamma_5$-scheme
independence of $a_i$, we take the NLO Wilson coefficients from
\cite{Buras96} (see Table 1). It is ready to check the scheme
dependence of $a_i$ from Eq. (\ref{ai}).

\begin{table}[ht]
\caption{Lowest-order (LO) and next-to-leading-order Wilson
coefficients in NDR and HV $\gamma_5$-schemes at $\mu=\overline{
m}_b(m_b)=4.40$ GeV for $\Lambda^{(5)}_{\overline{\rm MS}}=225$
MeV taken from Table XXII of [13], where $\alpha$ is the
fine-structure constant.}
\begin{center}
\begin{ruledtabular}
\begin{tabular}{ l c c c c c c c c c c}
 & $c_1$ & $c_2$ & $c_3$ & $c_4$
 & $c_5$ & $c_6$ & $c_7/\alpha$ & $c_8/\alpha$ & $c_9/\alpha$ &
 $c_{10}/\alpha$
 \\ \hline
 LO & 1.144 & -0.308 & 0.014 & -0.030 & 0.009 & -0.038 & 0.045 &
 0.048 & -1.280 & 0.328 \\
 NDR & 1.082 & -0.185 & 0.014 & -0.035 & 0.009 & -0.041 & -0.002 &
 0.054 & -1.292 & 0.263 \\
 HV & 1.105 & -0.228 & 0.013 & -0.029 & 0.009 & -0.033 & 0.005 &
 0.060 & -1.283 & 0.266 \\
\end{tabular}
\end{ruledtabular}
\end{center}
\end{table}

From Eq. (\ref{naivef}) we see that the $B\to\J K$ amplitude is
governed by the parameter
\be
\bar a_2\equiv a_2+a_3+a_5+a_7+a_9.
\en
In naive factorization, it is predicted to be $\bar a^{\rm
LO}_2(\J K)=0.074 $ using the leading-order Wilson coefficients at
$\mu=m_b$ given in Table I. Evidently it is too small compared to
the experimental value
\be
|\bar a_2(\J K)|_{\rm expt}=0.26\pm0.02,
\en
extracted from the data \cite{PDG}
\be
{\cal B}(B^-\to \J K^-)=(10.0\pm 1.0)\times 10^{-4}, \quad {\cal
B}(B^0\to \J K^0)=(8.9\pm 1.2)\times 10^{-4}.
\en

In ensuing calculations we will use the form factors
\be
F_1^{BK}(m_\J^2)= 0.70, \quad\qquad F_0^{BK}(m_\J^2)=0.50,
\en
from \cite{a12}, the decay constants $f_K=0.16$ GeV, $f_B=0.19$
GeV, $\Lqcd=300$ MeV, and the running quark masses \cite{Buras96}:
$m_b=4.40$ GeV and $m_c=1.30$ GeV. For the parameter $r$ in
(\ref{logdiv}), in principle it may be complex due to soft
rescattering. In \cite{BBNS2}, $r$ is chosen randomly inside a
circle in the complex plane of radius 3 (``realistic") or 6
(``conservative").

In deriving the hard scattering functions $f_I$, $g_I$,
$f_{II}^2$, $f_{II}^3$, we have neglected the difference between
$m_b$ and $m_B$ in the heavy quark limit. However, since
$f^3_{II}$ is sensitive to the value of $z$, we will vary $z$ from
one extreme $z=m_\J^2/m_B^2$ to the other extreme
$z=m_\J^2/m_b^2$, as shown in Table II, where we have used $r=4.5$
for illustration. The sensitivity of $f_{II}^3$ on $z$ is not
unexpected because the twist-3 effects on spectator interactions
are governed by soft gluon exchange. In Table II we show the
numerical values of the parameters  $a_2(\J K)$ and $\bar a_2(\J
K)$, and, for comparison, the twist-2 results of $\bar a_2(\J K)$
denoted by $\bar a^{\rm t2}_2(\J K)$, i.e. the predictions without
the twist-3 effect $f_{II}^3$.

Since {\it a priori} the shape of the $\J$ LCDA is unknown, it is
worth considering other possibilities besides the asymptotic form
which we have used thus far. The delta-function form
$\phi^\J(\xi)=\delta(\xi-{1\over 2})$, as suggested in
\cite{Chay}, appeals to the naive expectation of the wave
function. The results are shown in the last row in Table II. We
see that while $f_I$ and $g_I$ are not sensitive to the change of
the $\J$ wave function, the values of $f_{II}^2$ and $f_{II}^3$
are reduced by a factor of 2/3 owing to the fact that the integral
$\int^1_0(d\xi/\xi)\phi^\J(\xi)$ [see Eq. (\ref{fIIt})] is equal
to 3 for the asymptotic form and 2 for the delta form of the $\J$
wave function. As a consequence, the resultant $|\bar a_2(\J K)|$
becomes smaller.

\begin{table}[ht]
\caption{Numerical results of hard scattering functions and the
parameters $\bar a^{\rm t2}_2(\J K)$, $a_2(\J K)$, $\bar a_2(\J
K)$ for two choices of $z$ and for $r=4.5$. For comparison, the
predictions using the delta-function form for the $\J$
distribution amplitude are shown in the last row.}
\begin{center}
\begin{ruledtabular}
\begin{tabular}{ l c c c c c c c }
 & $f_I$ & $g_I$ & $f_{II}^2$ & $f_{II}^3$ & $\bar a^{\rm t2}_2(\J K)$ &
 $a_2(\J K)$ & $\bar a_2(\J K)$
 \\ \hline
 $z=m_\J^2/m_B^2$
& -1.17-i6.14 & 0.54-i0.74 & 4.89 & 12.71 & 0.051-i0.057 &
0.166-i0.056 & 0.158-i0.057  \\ $z=m_\J^2/m_b^2$ &  -1.08-i4.60 &
0.73-i1.27 & 6.36 & 21.47 & 0.065-i0.047 & 0.254-i0.046 &
0.247-i0.047
\\ \hline $z=m_\J^2/m_B^2$ & -1.03-i6.43 & 0.77-i0.79 & 3.26 &
8.47 & 0.040-i0.059 & 0.119-i0.059 & 0.111-i0.059  \\
\end{tabular}
\end{ruledtabular}
\end{center}
\end{table}

From Table II we see that at twist-2 order, $|\bar a^{\rm t2}_2(\J
K)|$ is of order $0.07\sim 0.08$ which is very close to the naive
prediction $a^{\rm LO}_2(\J K)=0.074$. This means that
nonfactorizable corrections in the heavy quark limit to naive
factorization is small. Therefore, the predicted branching ratio
of $B\to\J K$ to the leading-twist order is too small by a factor
of $7\sim 10$. When the unknown parameter $r=|r|\exp(i\delta)$ is
varied from 3 to 6 in the complex plane and $z$ varies from
$m_\J^2/m_B^2$ to $m_\J^2/m_b^2$, we find that $|a_2(\J K)|$ and
$|\bar a_2(\J K)|$ fall into the range $0.12\sim 0.29$ for
$\delta\leq 45^\circ$ (see Fig. 2), which is a reasonable choice
since the phase of the spectator diagrams is expected to be small.

\begin{figure}[t!]
 \begin{center}
\includegraphics[width=0.5\textwidth]{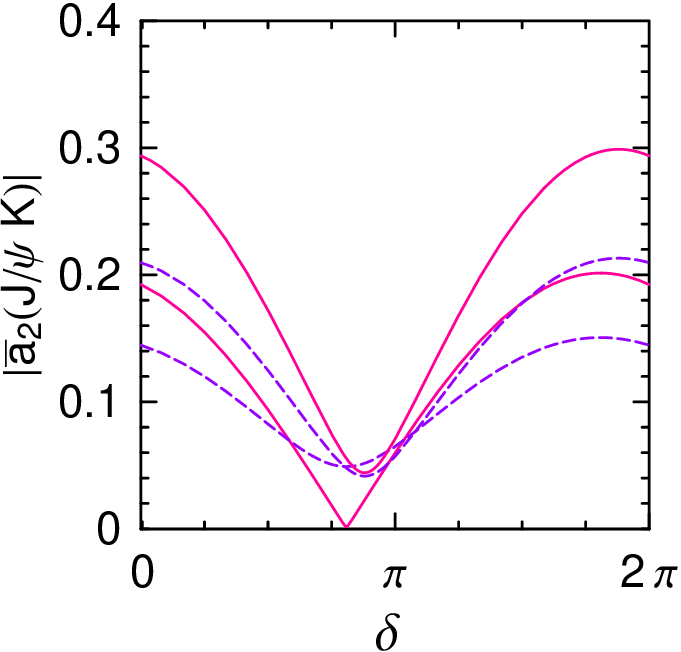}
\vspace{0.2cm}
    \caption{{\small The coefficient $|\bar a_2(\J K)|$ vs. the
    phase of the parameter $r$. Solid and dashed curves are for
    $|r|=6$ and 3, respectively.
    The upper and lower solid curves are for $z=m_\J^2/m_b^2$ and
    $m_\J^2/m_B^2$,
    respectively, and likewise for the dashed curves.}}
   \label{fig:2}
  \end{center}
\end{figure}

Needless to say, the unknown parameter $X$, the dependence of
$f_{II}^3$ on $z$ and the unknown $\J$ LCDAs are the main sources
of theoretical uncertainties. Hence, our result for $a_2(\J K)$
should be regarded as an order of magnitude estimate.
Nevertheless, it is clear that the discrepancy between theory and
experiment is greatly improved by the inclusion of kaon twist-3
effects and the sign of Re$a_2(\J K)$ is positive. It remains to
study the higher-twist effect of $\J$.

\section{Conclusions}

We have studied the decay $B\to\J K$ within the framework of the
QCD factorization approach. Physically, the factorization method
is still applicable because the transverse size of $\J$ becomes
small in the heavy quark limit so that its overlap with the $(BK)$
system is small. Technically, the infrared divergence due to soft
gluon exchange between the $c$ quark and the $(BK)$ system is
canceled by that between the $\bar c$ quark and $(BK)$ so that
nonfactorizable corrections are infrared safe. We have shown
explicitly that this is indeed the case to the twist-2 order.

The scale and $\gamma_5$-scheme problems with naive factorization
are resolved when radiative vertex corrections to the hadronic
matrix elements are included, as we have proved explicitly.
However, nonfactorizable corrections to naive factorization due to
vertex diagrams and hard spectator interactions are small.
Therefore, the predicted branching ratio of $B\to\J K$ to the
leading-twist order is too small by a factor of $7\sim 10$
compared to experiment.

Since the virtuality of the hard exchanged gluon in the spectator
diagrams is only of order $m_b\Lqcd$, it is expected that
power-suppressed terms arising from higher-twist wave functions
become important. We found that the contribution from the twist-3
kaon light-cone distribution amplitude $\phi^K_\sigma$ to
spectator interactions is not only chirally but also
logarithmically and kinematically enhanced, indicating that soft
gluon exchange between the spectator quark of the $B$ meson and
the charmed or anticharmed quark of the charmonium dominates.
Consequently, factorization breaks down at twist-3 level. In spite
of many theoretical uncertainties, the predicted $|a_2(\J K)|$ can
range from 0.12 to 0.29. This also solves the long-standing sign
ambiguity for Re$a_2(\J K)$. Though it remains to check the
higher-twist effects of the charmonium, it is expected that
twist-3 corrections which manifest predominately in hard spectator
interactions are dominated by the kaon.

\vskip 1.0cm \acknowledgments We are grateful to J. Chay for
interesting discussion. One of us (H.Y.C.) wishes to thank the
hospitality of the Physics Department, Brookhaven National
Laboratory. This work was supported in part by the National
Science Council of R.O.C. under Grant Nos. NSC89-2112-M-001-082
and NSC89-2112-M-033-014.


\end{document}